\documentclass[epjH]{svjour}
\usepackage{graphics}
\usepackage{bm}
\usepackage{amssymb}
\begin{document}
\title{Lorenz's electromagnetic theory of light}
\author{C. W. Wong\thanks{\email{cwong@physics.ucla.edu} } }
%
\institute{Department of Physics and Astronomy, 
University of California, Los Angeles, CA 90095-1547, USA}
\abstract{ The Lorenz electromagnetic theory of light, published two years after
the Maxwell theory, starts by postulating that both scalar and vector potentials
are retarded. We show that in spite of this postulate, Lorenz's theory gives a 
longitudinal electric field in vacuum that remains the instantaneous action at a 
distance it is in the Maxwell theory. There is in fact no difference between the 
two theories for electromagnetic phenomena in vacuum. 
} 
\maketitle
\section{Introduction}
\label{intro}

The retarded scalar potential found in almost every textbook of electromagnetic 
theory today was first presented by Riemann to the G\"ottingen Academy 
in 1858, but the short paper was subsequently withdrawn 
(footnote 2, \cite{Whittaker51}) \footnote{Riemann was reluctant to publish 
incomplete work. On his untimely death at age 40, he left much work unpublished.}. 
It was finally published posthumously in 1867 \cite{Riemann67}, a year after 
Riemann's death. With this retarded potential, Riemann proposed that the electrical 
action was not instantaneous, but propagated with light speed $c$.

In 1867, Lorenz \cite{Lorenz67} independently proposed that Kirchhoff's equations 
of electrical currents in conductors be slightly modified by requiring that these 
currents propagated with light speed in order to obtain an electrical theory of 
light. Drawing upon his previous results on wave propagation in elastic bodies
\cite{Lorenz61}, he proposed an electromagnetic theory of light even in a 
non-conducting space such as the vacuum by postulating that both scalar and 
vector potentials were really retarded and satisfied wave equations where light 
speed $c$ appeared. 

In this way, Lorenz introduced what is known today as the Lorenz gauge. For many
years, the gauge was mistakenly called the Lorentz gauge, after Lorentz who also 
used retarded potentials in a long paper in 1892 \cite{Lorentz92}. Fortunately, 
the story of the misattribution has recently been told in detail by Jackson and 
Okun \cite{Jackson01}. The misnaming of the Lorenz gauge is not the concern of 
this paper.

In this paper, we are interested rather in the differences between the Maxwell
and Lorenz theories of electromagnetism in vacuum. In his 1873 {\it Treatise}, 
Maxwell acknowledged that Lorenz's theory of light was `similar to' his own 1865 
theory \cite{Maxwell65} of the electromagnetic field, but considered Lorenz's work 
too little and too late ({\cite{Maxwell73}, p.183 \cite{ORahilly38}). To this day, 
there are Maxwellians who take the same point of view. 
However, there were \cite{ORahilly38} and still are \cite{Jackson01,Jackson10} 
Lorenzians who take Lorenz's retarded potentials literally and consider the 
Lorenz theory as the more correct description.

The scientific proposals of Maxwell and Lorenz to be discussed in this paper were
made almost 150 years ago when the mechanistic nature of the vacuum was a scientific
question of great interest. The term Maxwellian today can have a meaning very different 
from its true historical meaning. Our view of the physical world has changed
very substantially a century after J. J. Thompson's discovery of the electron in 
1897 and Einstein's 1905 invention of the special theory of relativity and analysis 
of the electrodynamics of moving bodies. Maxwell's vacuum equations have stood 
the test of time, however\footnote{Maxwell's equations contained the scalar and 
vector potentials defined in the Coulomb gauge. The more compact version given in 
many modern-day textbooks is the gauge-independent form written entirely in 
terms of the electromagnetic fields. This simplified version of the Maxwell 
equations was obtained independently by Heaviside in 1885 and by Hertz in 1884 
\cite{Nahin88}.}. In our modern view, Maxwell's displacement 
current is no longer the artificial and awkward device found so distasteful by the 
anti-Maxwellians. It appears instead as a natural partner to Faraday's induction 
in the electric-magnetic duality structure of electromagnetism. It is conceptually 
very satisfying that electromagnetic wave motion is another manifestation of the 
experimentally well established and technologically highly important phenomenon of 
Faraday induction.  

In this paper, we mean by a Maxwellian approach the formulation of the basic laws
of electromagnetism in terms of Faraday induction and Maxwell displacement. 
This approach is in stark contrast to Lorenz's proposal to start instead 
with a postulated retarded form of the scalar and vector potentials. Lorenz's 
justification was that the light speed $c$ though large was known experimentally to 
be finite. So he believed that the classical concept of action at a distance must 
be an idealization and an approximation.

The purpose of this paper is to determine how similar Lorenz's electromagnetic 
(EM) theory is to Maxwell's. We shall do this with the method of Fourier 
transform used previously in our study of the Maxwell theory 
\cite{Wong09,Wong10,Wong10a}. In this method, partial differential equations 
whose derivative terms carry constant coefficients are transformed into simple 
algebraic equations in Fourier space. It then becomes clear that the two theories 
agree in their treatments of the causal transverse EM fields and transverse vector 
potential. The apparent difference appears only in their scalar and longitudinal 
vector potentials. We then show that Lorenz's 
retarded scalar/vector potentials appear in the longitudinal electric field in a 
certain combination that makes their retardations cancel completely. What is left 
is the same action at a distance described by the time-dependent Coulomb/Gauss 
law of the Maxwell theory. Hence the two theories describe {\it exactly the same} 
phenomena from two different starting points. Neither Riemann nor Lorenz was 
correct in supposing that instantaneous action at a distance can be eliminated 
completely from electromagnetic phenomena.

\section{Lorenz's theory in Fourier space}
\label{sec:Lorenz}

In the Lorenz theory (\cite{Lorenz67}, pp. 267--270, \cite{Whittaker51}, chap. VI, 
pp. 181--202 \cite{ORahilly38}), one begins by postulating that the scalar and vector 
potentials satisfy the wave equations.
\begin{eqnarray}
\left( \nabla^2 - \frac{1}{c^2}\partial_t^2 \right) \Phi({\bf r}, t)
&=& - \frac{\rho({\bf r}, t)}{\epsilon_0 },
\nonumber \\
\left( \nabla^2 - \frac{1}{c^2}\partial_t^2 \right) {\bf A}({\bf r}, t)
&=& - \mu_0 {\bf J}({\bf r}, t),
\label{waveEqPhiA}
\end{eqnarray}
in SI units in the notation of Jackson \cite{Jackson99}. Here 
$\partial_t = \partial/\partial t$. We shall work in the Fourier space 
$({\bf k}, \omega)$ where all differential equations with constant coefficients 
become algebraic equations that can be manipulated transparently. 

Suppose these potentials  and their first to third derivatives in space-time 
have Fourier representations of the type
\begin{eqnarray}
\Phi({\bf r}, t) &=& \int_{-\infty}^\infty \frac{d\omega}{2\pi} 
\int \frac{d^3 k}{(2\pi )^3} e^{i{\bf k}\bm{\cdot}{\bf r} - i\omega t} 
\tilde{\Phi}({\bf k}, \omega), 
\label{FR-rho} \\
\bm {\nabla}\Phi({\bf r}, t) 
&=& \int_{-\infty}^\infty \frac{d\omega}{2\pi} 
\int \frac{d^3 k}{(2\pi )^3} e^{i{\bf k}\bm{\cdot}{\bf r} - i\omega t} 
i{\bf k}\tilde{\Phi}({\bf k}, \omega), 
\label{FR-gradRho} 
\end{eqnarray}
etc. Note that the differential operator $\bm{\nabla}$ simply operates on the
Fourier basis functions $\exp{(i{\bf k}\bm{\cdot}{\bf r} - i\omega t)}$. The wave 
Eqs. (\ref{waveEqPhiA}) then simplify to the algebraic equations in Fourier space 
\begin{eqnarray}
\left (k^2 - \frac{\omega^2}{c^2} \right) \tilde{\Phi}
&=&  \frac{\tilde{\rho}}{\epsilon_0 },
\nonumber \\
\left (k^2 - \frac{\omega^2}{c^2} \right) \tilde{\bf A} 
&=& \mu_0 \tilde{\bf J}. 
\label{waveEqsL-FS}
\end{eqnarray}

Since the operations $\bm{\nabla \cdot}$ and $\bm{\nabla \times}$ appear in 
Fourier space as the simple vector operations $i{\bf k}\bm{\cdot}$ and 
$i{\bf k}\bm{\times}$, the Helmholtz decomposition \cite{Helmholtz58} of a vector 
field in Fourier space simplifies to the BAC identity
\begin{eqnarray}
\tilde{\bf A} = \tilde{\bf A}_\parallel + \tilde{\bf A}_\perp
= {\bf e}_{\bf k} ({\bf e}_{\bf k} \bm{\cdot} \tilde{\bf A}) 
- {\bf e}_{\bf k} \bm{\times} ({\bf e}_{\bf k} \bm{\times} \tilde{\bf A}), 
\label{HelmholtzThm}
\end{eqnarray}
where ${\bf e}_{\bf k} = {\bf k}/k$. It gives a unique separation of the 
longitudinal and transverse (L/T) parts of the vector field $\tilde{\bf A}$. 
We can now write down all the remaining equations of the Lorenz theory in Fourier 
space without further ado. 

The electric field is defined to be
\begin{eqnarray}
\tilde{\bf E} =  - i{\bf k} \tilde{\Phi} + i\omega \tilde{\bf A}. 
\label{E-kw}
\end{eqnarray}
Its L/T parts are found by inspection:
\begin{eqnarray}
\tilde{\bf E}_\parallel &=&  - i{\bf k} \tilde{\Phi} 
+ i\omega \tilde{\bf A}_\parallel,
\label{EL} \\
\tilde{\bf E}_\perp &=&  i\omega \tilde{\bf A}_\perp. 
\label{ET}
\end{eqnarray}
The magnetic induction, defined to be
\begin{eqnarray}
\tilde{\bf B} =  i{\bf k} \bm{\times} \tilde{\bf A} 
= i{\bf k} \bm{\times} \tilde{\bf A}_\perp, 
\label{B-kw}
\end{eqnarray}
turns out to be purely transverse. Its vanishing longitudinal part
$\tilde{\bf B}_\parallel = 0$ describes the fact that no magnetic charges or
monopoles are present. We thus see that the same transverse causal vector
potential $\tilde{\bf A}_\perp$ determines both transverse causal EM fields 
uniquely. These results are the same as the Maxwell results. Indeed it had been 
realized by Young in 1817 (\cite{Peacock55}, pp. 114-5 \cite{Whittaker51}) 
and by Frenel in 1821 (\cite{Fresnel21}, pp. 115-22, \cite{Whittaker51}) that 
the two independent polarization states seen in light passing through crystalline 
solids could be understood only if light propagation was caused by transverse
vibrations and not longitudinal vibrations.

It is now useful to describe the consequence of the continuity 
equation for a conserved charge density:
\begin{eqnarray}
\frac{d\rho}{dt} = \partial_t\rho + \bm{\nabla \cdot}{\bf J} = 0
\label{ContinuityEq-rt} 
\end{eqnarray}
in spacetime, and  
\begin{eqnarray}
\tilde{J}_\parallel &=&  \frac{\omega}{k} \tilde{\rho}.
\label{ContinuityEq} 
\end{eqnarray}
in Fourier space. Used with the wave Eqs. (\ref{waveEqsL-FS}), it yields 
the expression
\begin{eqnarray}
\left( k^2 - \frac{\omega^2}{c^2} \right) 
\left( \epsilon_0\omega \tilde{\Phi} - \frac{k}{\mu_0}\tilde{\bf A} \right)
&=&  \omega\tilde{\rho} - k \tilde{J}_\parallel = 0.
\label{LorenzCondWaveEq}
\end{eqnarray}
This means that the Lorenz gauge condition
\begin{eqnarray}
\epsilon_0\omega \tilde{\Phi} - \frac{k}{\mu_0}\tilde{\bf A} = 0
\label{LorenzCond}
\end{eqnarray}
holds if $k^2 \neq \omega^2/c^2$. The inequality for $k^2$ appears because the Lorenz 
condition actually involves the solutions of two inhomogeneous wave equations whose 
sources when thus combined cancel everywhere in Fourier space or in spacetime. Thus 
the Lorenz theory seems to have the nice feature that the Lorenz condition comes 
out naturally from the postulated retarded scalar/vector potentials. However, its 
deeper significance is that this postulated retardation actually disappears 
completely from $\tilde{\bf E}_\parallel$, as we shall show in the following.

For completeness, we should mention that the Lorenz condition is violated for the 
solution of the homogeneous wave Eq. (\ref{LorenzCondWaveEq}) satisfying the 
condition $k^2 = \omega^2/c^2$, but the resulting complementary function is only 
concerned with a change of the boundary/initial conditions satisfied by a solution 
of an inhomogeneous wave equation.

We turn finally to the single scalar field in $\tilde{\bf E}_\parallel$. It can 
be expressed in terms of two of the solutions $\tilde{\rho}$ and 
$\tilde{A}_\parallel$ of the wave Eqs. (\ref{waveEqsL-FS}):
\begin{eqnarray}
\tilde{E}_\parallel &=&  - \frac{i}{\epsilon_0 k} 
\left(\frac{\tilde{\rho}k^2 - \tilde{J}_\parallel\omega k/c^2}{k^2 - \omega^2/c^2}
\right).
\label{EL2} 
\end{eqnarray}
Further simplification obtains on using the continuity Eq. 
(\ref{ContinuityEq}): 
\begin{eqnarray}
\tilde{E}_\parallel = - \frac{i\tilde{\rho}}{\epsilon_0 k} 
\left(\frac{k^2 - \omega^2/c^2}{k^2 - \omega^2/c^2} \right)
= - \frac{i\tilde{\rho}}{\epsilon_0 k}.
\label{EL3} 
\end{eqnarray}
The final step is justified because $k^2 \neq \omega^2/c^2$. Hence the Lorenz 
$\tilde{\bf E}_\parallel$ is no longer retarded, because it does not depend on 
the light speed $c$ anymore. In fact, it is just the Coulomb/Gauss law in the 
Maxwell equation 
\begin{eqnarray}
\bm{\nabla \cdot}{\bf E}({\bf r}, t) = \rho({\bf r}, t)/\epsilon_0
\label{Coulomb/Gauss} 
\end{eqnarray}
in spacetime. The div operator on the left introduces a spatial nonlocality,
but the action occurs at the same time $t$ on both sides of the equation.  
${\bf E}_\parallel$ is therefore also an instantaneous action at a distance
in the Lorenz theory. 

In conclusion, we have shown that the Lorentz theory gives exactly the same 
EM fields as the Maxwell theory. In particular, ${\bf E}_\parallel$ remains an
instantaneous action at a distance because the retarded contributions to 
${\bf E}_\parallel$ from $\rho$ and ${\bf J}_\parallel$ cancel completely.

The Maxwell and Lorenz theories are two examples of a common 
gauge-invariant theory with the universal characteristics that all causal EM 
fields are transverse and that ${\bf E}_\parallel$ is an instantaneous action
at a distance \cite{Wong09,Wong10a}. The Maxwell and Lorenz theories are special 
cases of the common universal theory realized in the Coulomb and Lorenz gauges, 
respectively. More specifically, the result that ${\bf E}_\parallel$ is non-causal 
and not associated with light propagation is obtained in Lorenz theory with the 
help of the continuity equation for the charge density and in the Maxwell theory 
directly from the Coulomb/Gauss law.

\end{document}